\documentclass[prc,twocolumn,showpacs,preprintnumbers,amsmath,amssymb]{revtex4}

\usepackage{graphicx}
\usepackage{dcolumn}
\usepackage{bm}
\usepackage{CJK}
\usepackage{tipa}
\usepackage{amssymb}
\usepackage{booktabs}

\begin{document}
\begin{CJK*}{GBK}{song}

\title{The decay characteristic of $^{22}$Si and its ground-state mass significantly affected by three-nucleon forces}

\author{X.~X.~Xu$^{1}$}
\email[]{xuxinxing@ciae.ac.cn}
\author{C.~J.~Lin$^{1}$}
\email[]{cjlin@ciae.ac.cn}
\author{L.~J.~Sun$^{1}$}
\author{J.~S.~Wang$^{2}$}
\author{Y.~H.~Lam$^{2}$}
\author{J.~Lee$^{3}$}
\author{D.~Q.~Fang$^{4}$}
\author{Z.~H.~Li$^{5}$}
\author{N.~A.~Smirnova$^{6}$}
\author{C.~X.~Yuan$^{7}$}
\author{L.~Yang$^{1}$}
\author{Y.~T.~Wang$^{4}$}
\author{J.~Li$^{5}$}
\author{N.~R.~Ma$^{1}$}
\author{K.~Wang$^{4}$}
\author{H.~L.~Zang$^{5}$}
\author{H.~W.~Wang$^{4}$}
\author{C.~Li$^{4}$}
\author{M.~L.~Liu$^{2}$}
\author{J.~G.~Wang$^{2}$}
\author{C.~Z.~Shi$^{4}$}
\author{M.~W.~Nie$^{4}$}
\author{X.~F.~Li$^{4}$}
\author{H.~Li$^{4}$}
\author{J.~B.~Ma$^{2}$}
\author{P.~Ma$^{2}$}
\author{S.~L.~Jin$^{2}$}
\author{M.~R.~Huang$^{2}$}
\author{Z.~Bai$^{2}$}
\author{F.~Yang$^{1}$}
\author{H.~M.~Jia$^{1}$}
\author{Z.~H.~Liu$^{1}$}
\author{D.~X.~Wang$^{1}$}
\author{Y.~Y.~Yang$^{2}$}
\author{Y.~J.~Zhou$^{2}$}
\author{W.~H.~Ma$^{2}$}
\author{J.~Chen$^{2}$}
\author{Z.~G.~Hu$^{2}$}
\author{Y.~H.~Zhang$^{2}$}
\author{X.~W.~Ma$^{2}$}
\author{X.~H.~Zhou$^{2}$}
\author{Y.~G.~Ma$^{4}$}
\author{H.~S.~Xu$^{2}$}
\author{G.~Q.~Xiao$^{2}$}
\author{H.~Q.~Zhang$^{1}$}

\affiliation{$^1$Department of Nuclear Physics, China Institute of Atomic Energy, Beijing 102413, People's Republic of China\\
$^2$Institute of Modern Physics, Chinese Academy of Sciences, Lanzhou 730000, People's Republic of China\\
$^3$Department of Physics, The University of Hong Kong, Hong Kong, People's Republic of China\\
$^4$Shanghai Institute of Applied Physics, Chinese Academy of Sciences, Shanghai 201800, People's Republic of China\\
$^5$State Key Laboratory of Nuclear Physics and Technology, School of Physics, Peking University, Beijing 100871, People's Republic of China\\
$^6$CENBG, CNRS/IN2P3 and Universite de Bordeaux, Chemin du Solarium, 33175 Gradignan cedex, France\\
$^7$Sino-French Institute of Nuclear Engineering and Technology, Sun Yat-Sen University, Zhuhai 519082, China}

\date{\today}

\begin{abstract}
The decay of the proton-rich nucleus $^{22}$Si was studied by a silicon array coupled with germanium clover detectors. Nine charged-particle groups are observed and most of them are recognized as $\beta$-delayed proton emission. A charged-particle group at 5600 keV is identified experimentally as $\beta$-delayed two-proton emission from the isobaric analog state of $^{22}$Al. Another charged-particle emission without any $\beta$ particle at the low energy less than 300 keV is observed. The half-life of $^{22}$Si is determined as 27.5 (18) ms. The experimental results of $\beta$-decay of $^{22}$Si are compared and in nice agreement with shell-model calculations. The mass excess of the ground state of $^{22}$Si deduced from the experimental data shows that three-nucleon (3N) forces with repulsive contributions have significant effects on nuclei near the proton drip line.
\end{abstract}

\maketitle
\end{CJK*}

Impressive progresses in nuclear decay studies near the proton drip line have been achieved over recent decades ~\cite{rf01,rf02,rf03}. These exotic decay modes of proton-rich nuclei, such as latest studies of two-proton (2{\it p}) emission from ground states (e.g., $^{45}$Fe ~\cite{rf04}, $^{54}$Zn ~\cite{rf05}, and $^{30}$Ar ~\cite{rf06}) and excited levels ~\cite{rf07,rf08}, and $\beta$-delayed particle emission ($^{31}$Ar ~\cite{rf09,rf10} and $^{20}$Mg ~\cite{rf11}), play a significant role in studies of nuclear structure, quantum many-body systems, and nuclear astrophysics. Recently, three-nucleon forces have been used to investigate proton-rich nuclei ~\cite{rf12}, which will build an important bridge between nuclear forces and decay near the proton drip line.

The lightest nucleus with an isospin projection T$_{z}$=-3, $^{22}$Si, was discovered nearly thirty years ago in GANIL ~\cite{rf13}. Up to now, only one experiment ~\cite{rf14} has been performed to study its spectroscopic information, in which $\beta$-delayed proton emission was observed and the half-life of $^{22}$Si was determined.  The $\beta$-decay of $^{22}$Si is of particular importance as high-quality shell-model calculations can be performed for $1s$-$0d$-shell nuclei and comparisons of relevant theoretical and experimental results can be made in order to check the reliability of the shell-model near the proton drip line. $^{22}$Si is also a candidate for $\beta$-delayed 2{\it p} and 3{\it p} emissions via the isobaric analog state (IAS) ~\cite{rf15}, as well as possible $\beta$-delayed {\it p}$\alpha$ emission ~\cite{rf14}. Moreover, $^{22}$Si may be unbound with respect to 2{\it p} emission to the ground state of $^{20}$Mg according to the atomic mass evaluation ~\cite{rf16,rf17}. Therefore, a new experiment was carried out to further investigate exotic decay properties of $^{22}$Si.

The experiment was performed at the National Laboratory of Heavy Ion Research (HIRFL) of the Institute of Modern Physics, Lanzhou, China. $^{22}$Si was produced by projectile fragmentation of a primary $^{28}$Si beam, accelerated to 75.8 MeV/{\it u} by HIRFL cyclotrons, which impinged with an average intensity of 37 enA on a 1500 $\mu$m $^{9}$Be target. The fragments were separated and purified by the first Radioactive Ion Beam Line in Lanzhou (RIBLL1)~\cite{rf18}. Two plastic scintillators at the second and fourth focal planes of the RIBLL1 and a silicon array~\cite{rf19} coupled with germanium clover detectors at the end of RIBLL1 were used to identify secondary ions on an event-by-event basis and study their decay properties with an implantation-decay correlation.

In the silicon array, two thin double-sided silicon strip detectors (DSSDs) (149 $\mu$m and 66 $\mu$m, respectively) in the center of the array served to measure the residual energy of the fragments and their decay characteristics. The thinner one (DSSD2) was used to detect low-energy decay protons as $\beta$ particles make less contributions to the background of the decay spectrum in DSSD2. A quadrant silicon detector (QSD1) was placed behind DSSD2 to achieve anticoincidence with the penetrating fragments and detect high-energy protons escaping from DSSD2. After QSD1, a 1546 $\mu$m thick quadrant silicon detector (QSD2) was installed for $\beta$ measurements. In addition, four 1500 $\mu$m thick quadrant silicon detectors were mounted upstream around the beam to detect $\beta$ particles and protons escaping from DSSDs. In this arrangement, the silicon array has the advantage of high-detection efficiency for $\beta$ particles (43$\%$). The low-energy threshold of 150 keV was determined by two-dimensional energy-correlation spectrum for protons escaping from one DSSD and deposited in the other in the decay of high statistical $^{20}$Mg.

The spectra of charged-particle emission from $^{22}$Si measured by (a)DSSD1, (b)DSSD2 and (c) their sum are shown in Fig.1, in which black and red lines represent the spectra for full-energy protons and those in coincidence with $\beta$ particles detected by the downstream QSD2, respectively. Nine proton groups are identified in total and some of them were reported and recognized as the $\beta$-delayed proton emission in the previous work by Blank et al ~\cite{rf14}. With the lower detection threshold, higher statistics and anticoincidence with escaping protons, five new peaks are observed in the present work. Their center-of-mass energies with a resolution of 50 keV (70 keV for peak 9), absolute intensities and decay modes are displayed in Table~\ref{tab:table1}.  The absolute charged-particle branching ratio is determined as $101.3 (82) \%$ which is in good agreement with the result of Blank et al ~\cite{rf14}.

\begin{figure}
\includegraphics[width=3.5in,height=5.3in]{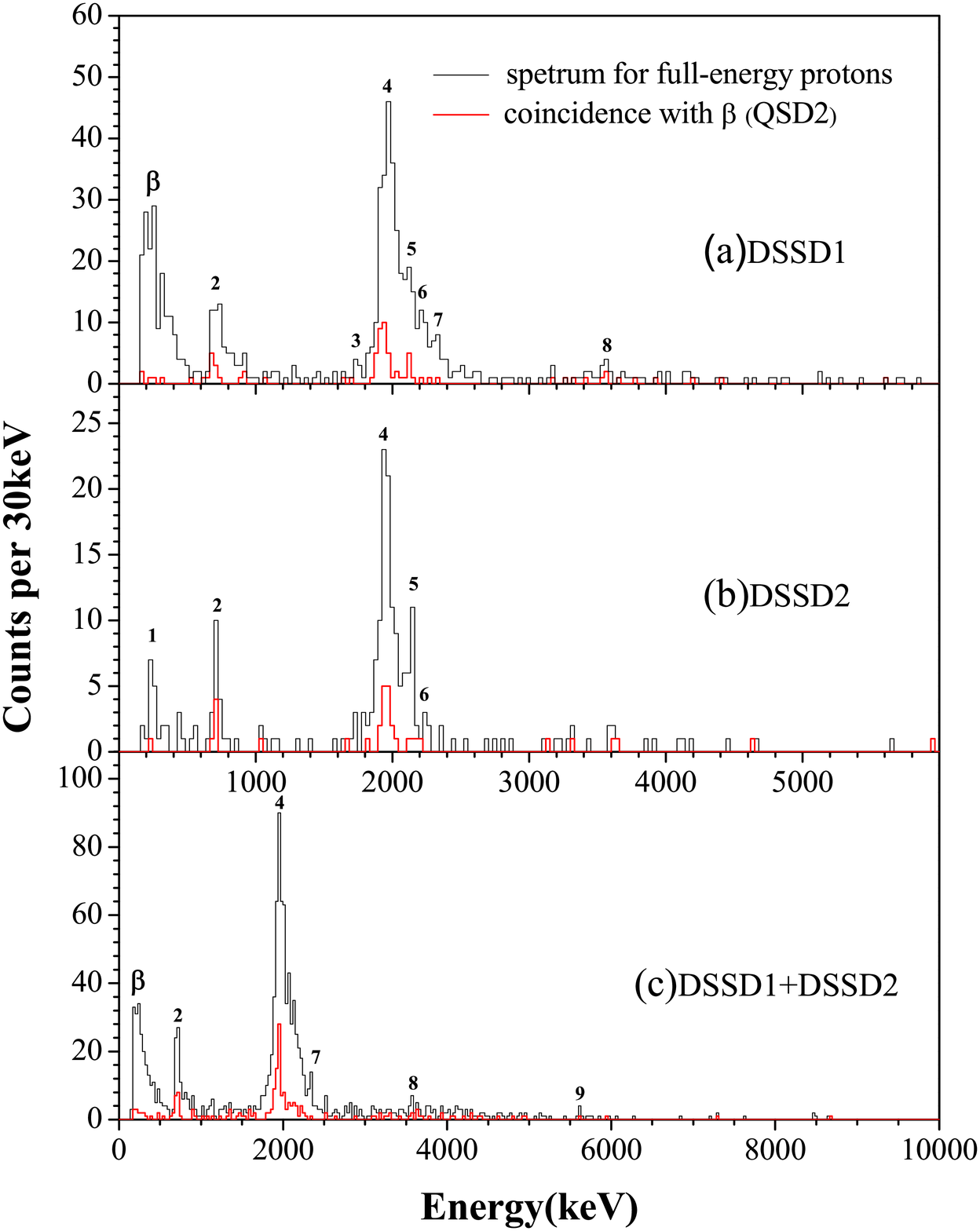}
\caption{\label{fig1} Charged-particle spectra in the decay of $^{22}$Si measured by (a)DSSD1 (149 $\mu$m), (b)DSSD2 (66 $\mu$m) and (c) their sum. Escaping protons from DSSDs can be largely rejected by surrounding detectors and only full-energy protons stopped in DSSDs are registered in spectra. The peak labels correspond to the peak numbers used in Table~\ref{tab:table1}.}
\end{figure}

\begin{table}
\caption{\label{tab:table1}
Center-of-mass energies, absolute intensities and decay modes for the identified charged-particle groups in the decay of $^{22}$Si. The energy resolution of these peaks was determined as about 50 keV (70 keV for peak 9).
}
\begin{ruledtabular}
\begin{tabular}{ccdc}
\textrm{Peak}&
\textrm{Energy (keV)}&
\multicolumn{1}{c}{\textrm{Absolute Intensity ($\%$)}}&
\textrm{Decay Mode}\\
\colrule
1 & 230 & 2.9 (10) &possible 2p$_{\it 0}$\\
2 & 680 & 6.8 (14) & $\beta$p$_{\it 3}$\\
3 & 1710 & 1.9 (7) & $\beta$p$_{\it 1}$\\
4 & 1950 & 52.0 (74) & $\beta$p$_{\it 1}$\\
5 & 2110 & 10.9 (21) & $\beta$p$_{\it 1}$\\
6 & 2180 & 6.5 (15)& $\beta$p$_{\it 0}$\\
7 & 2330 & 5.1 (13) & $\beta$p$_{\it 0}$\\
8 & 3550 & 2.5 (9) & $\beta$p$_{\it 1}$\\
9 & 5600 & 0.7 (3) & $\beta$2p$_{\it 0}$\\
\end{tabular}
\end{ruledtabular}
\end{table}

\begin{figure}
\includegraphics[width=3.3in,height=2.8in]{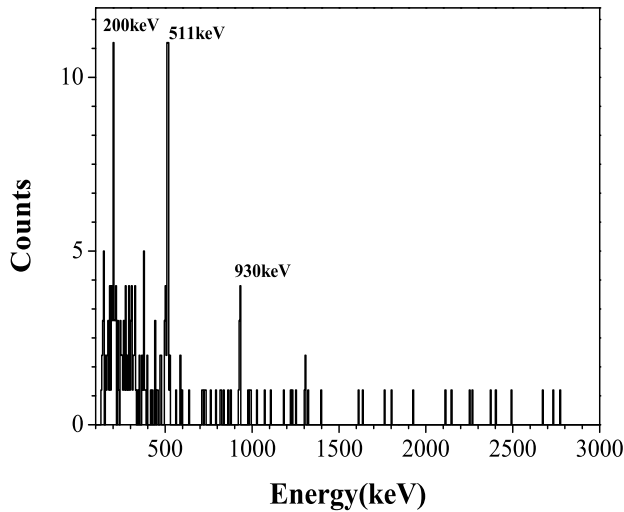}
\caption{\label{fig2} The $\gamma$-ray spectrum in the decay of $^{22}$Si. The $\gamma$ line observed at 200 keV corresponds to the de-excitation of the first excited state of $^{21}$Mg populated in the $\beta$p decay of $^{22}$Si.}
\end{figure}

The prominent proton group at 1950 (50) keV was identified to be the same transition as the one at 1990 (50) keV in the work of Blank et al ~\cite{rf14} based on its high intensity. It was attributed to an emission from an excited level of $^{22}$Al towards the first excited state of $^{21}$Mg, referred to as $\beta$p$_{\it 1}$ (4th column of Table~\ref{tab:table1}). To verify this assignment, a $\beta$-p-$\gamma$ coincidence was realized experimentally in the present work. Figure 2 shows the $\gamma$-ray spectrum obtained in the decay of $^{22}$Si. The $\gamma$ line at 200 keV is associated with the de-excitation of the first excited state in $^{21}$Mg towards its ground state and observed in coincidence with the proton group at 1950 keV. On this basis, the proton peak at 2180 keV was attributed to the decay of the same state [E$^{*}$= 2150 (52) keV, I$^{\pi}$=1$^+$] to the ground state of $^{21}$Mg, referred to as $\beta$p$_{\it 0}$ (6th column of Table~\ref{tab:table1}). Due to the quite low $\gamma$ detection efficiency and the weakness of most of proton transitions, other $\gamma$ transitions of excited states in $^{21}$Mg were not observed. The $\gamma$ line at 930 keV could not be assigned to any transition by means of a $\beta$-p-$\gamma$ coincidence nor any other method.

\begin{table}
\caption{\label{tab:table2}
Energy losses of measurements for the particles at the energy of 5600 keV and calculations for one proton with the same initial energy and path length in the implantation detector show that the particles at peak 9 cannot be one proton and must be two protons as the $\alpha$ particle with the energy of 5600 keV cannot escape from the implantation detector. Besides, two protons were obviously detected by different silicon strips for the events of number 2 and 4.}
\begin{ruledtabular}
\begin{tabular}{cccc}
\textrm{No.}&
\textrm{Path length ($\mu$m)}&
\textrm{Peak 9 (keV)}&
\textrm{One proton (keV)}\\
\colrule
1 & 32.2 (50) & 680 (50) & 420 (70)\\
2 & 35.7 (99) & 1930 (50) & 470 (130)\\
3 & 66.1 (177) & 3420 (50) & 900 (260)\\
4 & 49.3 (91) & 4980 (50) & 660 (130)\\
5 &  -   & 5600 (50) &   -\\
\end{tabular}
\end{ruledtabular}
\end{table}

Based on the consideration of the energy-level difference of $^{21}$Mg, the three transitions at 680 (50) keV, 2110 (50) keV and 2330 (50) keV could be assigned to the proton decay from the same excited state of $^{22}$Al [E$^{*}$= 2330 (52) keV, I$^{\pi}$=1$^+$] to the third excited, first excited and ground states of $^{21}$Mg, referred to as $\beta$p$_{\it 3}$, $\beta$p$_{\it 1}$, and $\beta$p$_{\it 0}$, respectively.

The proton groups at 1710 (50) keV and 3550 (50) keV were attributed to transitions from excited levels of 1910 keV and 3750 keV in $^{22}$Al to the first excited state of $^{21}$Mg, respectively, as the state of J$^{\pi}$ = 1/2$^{+}$ favors transitions from J$^{\pi}$ = 1$^{+}$ levels according to theoretical calculations. The two excited levels of $^{22}$Al may decay to other states of $^{21}$Mg, which will contribute to the distribution of Fig.1. However, due to very low intensities, the contribution was not taken into consideration in the analysis of branching ratios.

\begin{table*}
\caption{\label{tab:table3}
Properties of $\beta$-delayed charged-particle emission from $^{22}$Si. Experimental results are compared to theoretical calculations in which only $\beta$ decay is included and possible direct 2p emission is not taken into consideration in determining absolute branching ratios. The energies of excited states are deduced according to proton energies, the ground-state mass of $^{22}$Al and $^{21}$Mg ($^{20}$Na for the IAS) based on Nubase2012 ~\cite{rf17}, which uncertainties are only derived from those of proton energies and the mass of $^{21}$Mg ($^{20}$Na for the IAS).}
\begin{ruledtabular}
\begin{tabular}{lccccccccccc}
\multicolumn{4}{c}{\textrm{Excited state of $^{22}$Al (keV)}}&
\multicolumn{3}{c}{\textrm{Branching ratio ($\%$)}}&
\multicolumn{3}{c}{\textrm{log$ft$}}&
\multicolumn{1}{c}{\textrm{Decay mode}}&
\multicolumn{1}{c}{\textrm{Intensity ($\%$)}}\\
J$^{\pi}$ &Exp. & cd-USDB & wb-USD & Exp. & cd-USDB & wb-USD & Exp. & cd-USDB & wb-USD & & \\
\colrule
1$^{+}$ &1910 (52) & 1622 & 1300  & 1.9 (7) & 12.8 & 1.0 & 5.33 (17) & 4.58 & 5.81 & $\beta$p$_{\it 1}$ & 100 \\
1$^{+}$ &2150 (52) & 2533 & 2309  & 58.5 (76) & 54.3 & 56.8 & 3.79 (7)& 3.78 & 3.88 & $\beta$p$_{\it 0}$ & 11 (3) \\
 & &  &  &  &  &  & & & & $\beta$p$_{\it 1}$ & 89 (17) \\
1$^{+}$ &2330 (52) & 3299 & 2897 & 22.8 (28) & 5.3 & 22.3 & 4.17 (7)& 4.65 & 4.18 & $\beta$p$_{\it 0}$& 22 (6) \\
 & &  &  &  &  &  & & & & $\beta$p$_{\it 1}$ & 48 (11) \\
 & &  &  &  &  &  & & & & $\beta$p$_{\it 3}$ & 30 (7) \\
1$^{+}$ & - & 4334 & 3775  & - & 0.0 & 1.9 & - & 11.45 & 5.08 & - & - \\
1$^{+}$ &3750 (52) & 4831 & 3995 & $>$ 2.5 (9) & 7.6 & 10.3 & $<$ 4.83 (17)& 4.10 & 4.30 & $\beta$p$_{\it 1}$& $<$ 100 \\
0$^{+}$ \footnote{Isobaric Analog State of $^{22}$Al.} & 8830 (70) & 8576 & 9020 & $>$ 0.7 (3) & 10.7 & 7.7 & $<$ 3.75 (20) & 3.01 & 3.01 & $\beta$2p$_{\it 0}$& $<$ 100 \\
\end{tabular}
\end{ruledtabular}
\end{table*}

$^{22}$Si is also a candidate for $\beta$-delayed 2p emission via the isobaric analog state (IAS). According to the predictions based on systematics ~\cite{rf02}, the expected energy for a 2p decay of the IAS to the ground state in $^{20}$Na is about 5610 (220) keV. In the spectrum of the sum of DSSD1 and DSSD2, a charged-particle group at 5600 (70) keV with very low statistics of only 5 events has been observed obviously. To provide evidence for the decay mode, the decayed-light-particle identification with the energy-loss method was realized in the experiment. $^{22}$Si was implanted in one of DSSDs, then decayed and emitted charged particles which may escape from the implantation DSSD and be detected by the other DSSD. In this case, the path length of the emitted particles within the implantation DSSD can be deduced based on the kinematics of implantation and decay.

Table~\ref{tab:table2} shows the path lengths, measured energy losses for particles at 5600 keV in the implantation detector, and calculated energy losses for one proton with the same initial energy and path length. Energy losses of detected charged particles are larger than those of one proton indicating that the particle should be heavier. On the other hand, the range of $\alpha$ with the energy of 5600 keV is only 28 $\mu$m and it cannot escape from the implantation detector. Therefore, experimental results strongly suggest that the peak at 5600 keV belongs to $\beta$-delayed 2p emission. Moreover, for the event of number 2, two protons both escaped from the implantation DSSD and were detected by two different strips of the other DSSD. For the event of number 4, one proton was detected by the adjacent strip of the implantation one of the same DSSD, and the second proton escaped from the implantation DSSD and hit the other DSSD. In other words, two protons were observed experimentally for the events of number 2 and 4. The path lengths of two protons for the event number 5 cannot be deduced as two protons were both stopped in the implantation detector.

With the simulated detection efficiency of $70 (7) \%$ under the assumption of 2p emission in phase space, the experimental branching ratio of this peak is determined to be $0.7 (3) \%$ which is in accordance with the former experimental result (less than $1\%$) ~\cite{rf14}.

Experimental results of $\beta$-delayed charged-particle emission from $^{22}$Si are compared with shell-model (SM)  calculations of two different Hamiltonians in Table~\ref{tab:table3}. For cd-USDB, we performed calculations in the full $sd$ shell using an isospin non-conserving Hamiltonian ~\cite{rf20} composed by an isospin conserving Hamiltonian, i.e. USDB interaction~\cite{rf21}, a two-body Coulomb interaction adjusted with short range correlation scheme which is based on unitary correlation operator method (UCOM)~\cite{rf22}, a phenomenological charge-dependent part describing the isospin-symmetry breaking of the effective nucleon-nucleon interaction, and isovector single particle energies. Another Hamiltonian wb-USD ~\cite{rf23}, which is based on USD, considers the weakly bound effect of the proton $1s_{1/2}$ orbit contributed by Coulomb interaction. The monopole based universal interaction V$_{\text{MU}}$ ~\cite{rf24} plus M3Y spin-orbit force ~\cite{rf25} are used to calculate the two-body matrix elements (TBME) differences between the proton-proton and neutron-neutron terms involving $1s_{1/2}$ orbit. The validity of the V$_{\text{MU}}$ plus M3Y spin-orbit force is also examined in the neutron-rich nuclei in the $psd$ region ~\cite{rf26}. The  Hamiltonian cd-USDB concentrates on the isospin asymmetry effect in the interaction, while the wb-USD focuses on the weakly bound effect on the wave function of proton $1s_{1/2}$ orbit originated from the Coulomb interaction.

Table~\ref{tab:table3} presents shell-model results of the first five $1^{+}$ states and the IAS $0^{+}$ in $^{22}$Al. Both two SM results agree well with the observed data. The wb-USD gives less excitation energies than cd-USDB for all five $1^{+}$ states, indicating the weakly bound effect on the levels. In terms of the branching ratio of the decay of the $0^+_\textnormal{IAS}$ state, it is shown in theoretical calculations that the width of proton emission (see Refs.~\cite{rf27,rf28} for similar calculations) is much larger than the one of $\gamma$ decay. Therefore, there is a very minimal impact on the Gamow-Teller strength evaluation due to the $\beta$-delayed $\gamma$-proton decay ~\cite{rf29}. Possible $\beta$-delayed proton emission from the $0^+_\textnormal{IAS}$ state was unobserved because of the low detection efficiency of protons greater than 5 MeV for DSSDs and many branches of transitions to different excited states of $^{21}$Mg.

\begin{figure}
\includegraphics[width=3.3in,height=3.2in]{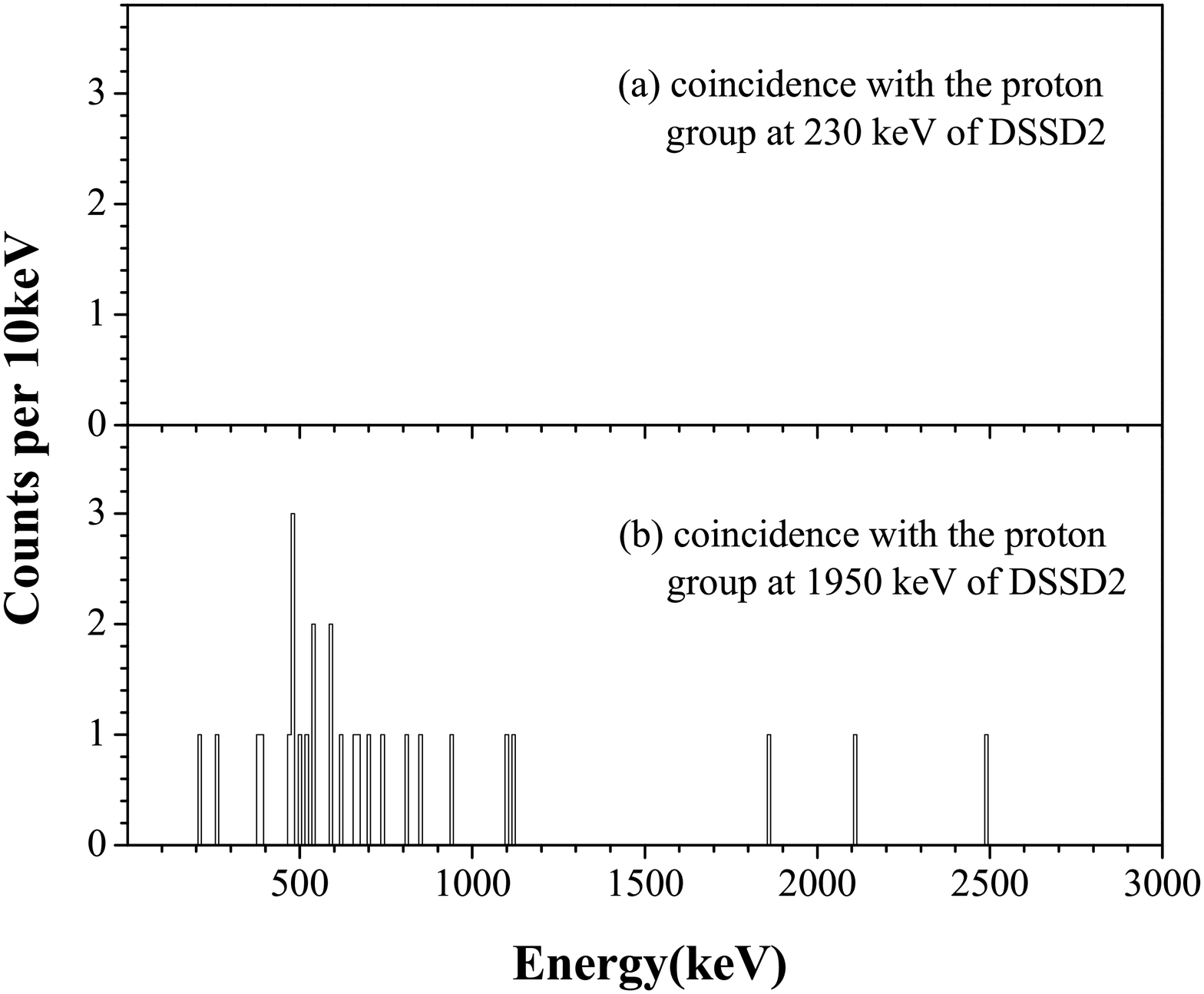}
\caption{\label{fig3} $\beta$-particle spectra in coincidence with the proton groups at (a) 230 keV and (b) 1950 keV of DSSD2, in which $\beta$ was measured by four 1500-$\mu$m-thick quadrant silicon detectors mounted upstream around the beam with the detection efficiency of $26\%$.}
\end{figure}

\begin{figure}
\includegraphics[width=3.6in,height=3.0in]{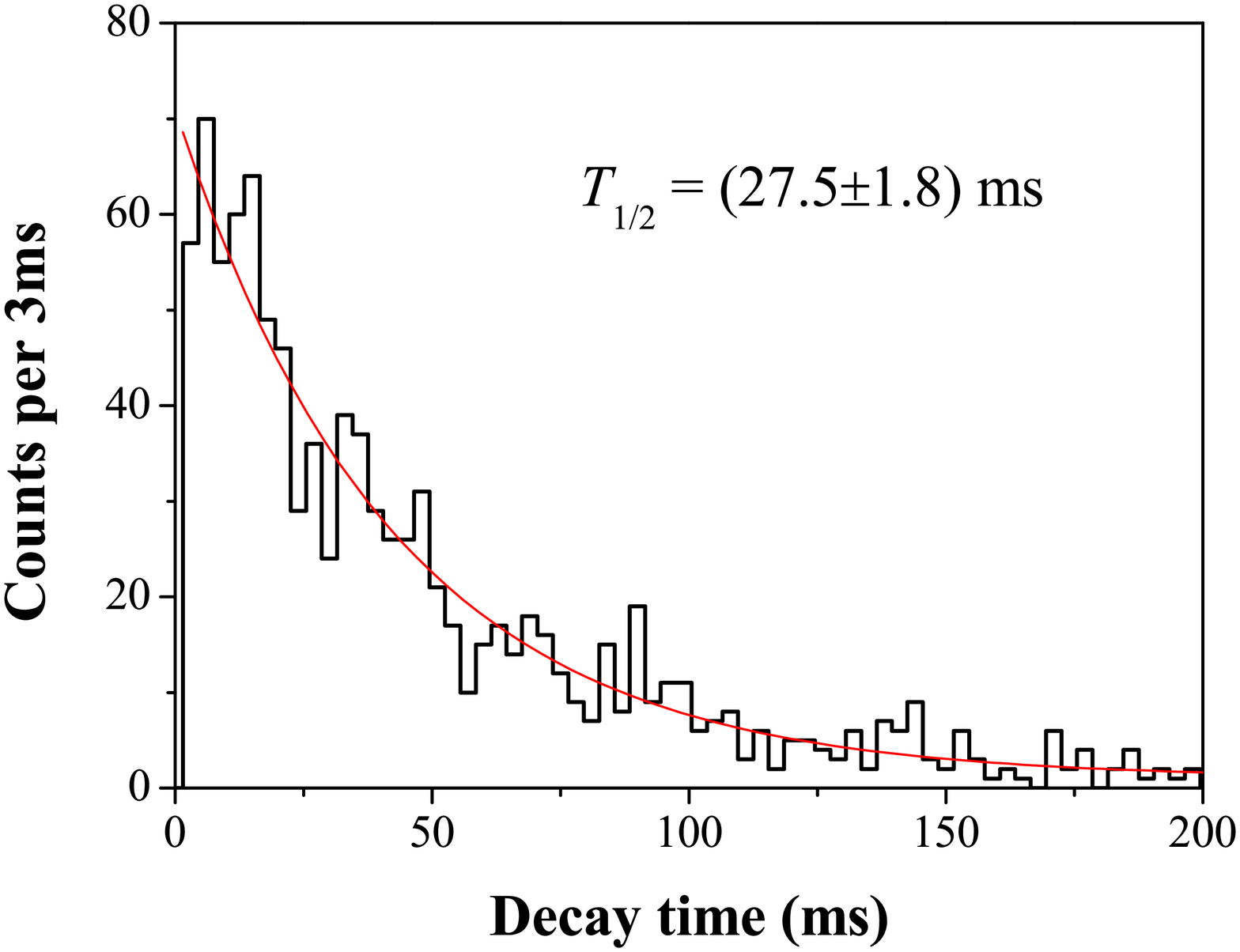}
\caption{\label{fig4} Decay-time spectrum of $^{22}$Si. A fit with a formula composed of an exponential decay and a constant background yields a half-life of 27.5 (18) ms for $^{22}$Si.}
\end{figure}

The charged-particle spectrum measured by DSSD2, Fig. 1 (b), exhibits a pronounced peak at 230 (50) keV. The events in this peak have no coincident $\beta$-particle signals in the four 1500-$\mu$m-thick quadrant silicon detectors mounted upstream around the beam, whereas the coincident $\beta$ particles can be observed for the events of the peak at 1950 keV with the same condition, as shown in Fig. 3. It should be pointed out that, an event at 230 keV in coincidence with a $\beta$ particle measured by QSD2 in Fig. 1 (b) may be an escaping $\beta$-delayed proton and cannot be a $\beta$ particle as its energy loss would be only about 20 keV if it was measured by the downstream QSD2. In this sense, the peak at 230 keV corresponds to a charged-particle emission without any $\beta$ particle and may be a direct 2p ground-state decay of $^{22}$Si. However, due to the poor statistics, the decay information of $^{20}$Mg, the 2p daughter nucleus, could not be obtained in the experiment. And because of the low decay energy, two protons cannot escape from DSSD2 and be directly identified by other detectors.

Figure 4 shows the decay-time spectrum of $^{22}$Si gated by decay energies less than 500 keV in Fig. 1(b) and greater than 500 keV in Fig. 1(c). The events of the gate up to 200 ms after $^{22}$Si implantation are fitted and the half-life thus obtained is 27.5 (18) ms, consistent with the previous experimental data of 29 (2) ms ~\cite{rf14}. Theoretical $\beta$-decay half-lives of $^{22}$Si determined by cd-USDB and wb-USD are estimated to be 29.5~ms and 21.0~ms, respectively.

The mass excess of the ground state of $^{22}$Si can be deduced from the equation: $\Delta$($^{22}$Si)= $\Delta$(IAS)+$\Delta$E$_{c}$-$\Delta$$_{nH}$ ~\cite{rf30}, where $\Delta$(IAS) is the mass excess of the IAS of $^{22}$Al and can be determined as 27029 (70) keV with the sum of the decay energy of $\beta$2p, the mass of two hydrogen atoms and the mass of $^{20}$Na ~\cite{rf17}. $\Delta$E$_{c}$ is the Coulomb energy difference between the ground state of $^{22}$Si and the IAS of $^{22}$Al, which can be deduced to be 5915.5 (90) keV. $\Delta$$_{nH}$ = 782.35 keV is the mass excess difference between a neutron and a hydrogen atom ~\cite{rf17}. In this way, the atomic mass excess of $^{22}$Si was deduced to be 32160 (115) keV. In another way, it is estimated to be 32367 (57) keV based on possible direct 2p decay observed experimentally. Two-proton separation energy (S$_{2p}$) for $^{22}$Si can be determined as -23 (115) keV and possible -230 (50) keV according to these two methods, respectively.

In calculations with 3N forces for ground-state energies of N = 8 isotones relative to the $^{16}$O core ~\cite{rf12}, $^{22}$Si is the most important case to show strong repulsive contributions of the order of several MeV originated from 3N forces as these forces in $^{22}$Si and $^{21}$Al have most significant effects and $^{21}$Al is difficult to be measured experimentally because of its very short half-life. Two-proton separation energy of $^{22}$Si obtained in our experiment is in very good agreement with the calculated value (S$_{2p}$ = -120 keV) of 3N forces in an extended sdf$_{7/2}$p$_{3/2}$ valence space, which show that 3N forces with repulsive contributions due to the Pauli exclusion principle~\cite{rf31} can play an important role in proton-rich nuclei and may have an affect on their decay characteristics near the proton drip line.

In summary, detailed studies devoted to the decay property and mass of the lightest nucleus with T$_{z}$=-3, $^{22}$Si, are presented in this letter. Firstly, new $\beta$-delayed 1p and 2p emissions have been identified obviously. The half-life and absolute decay branching ratios by counting implantation ions and decays are determined experimentally, which are in good agreement with shell-model calculations. Secondly, a charged-particle emission without any $\beta$ particle at the very low energy less than 300 keV is observed and may be a direct 2p ground-state decay of $^{22}$Si. Thirdly, the mass excess of the ground state of $^{22}$Si is estimated according to $\beta$-delayed 2p emission and possible direct 2p emission, respectively, and agrees well with the recent calculation which shows strong repulsive contributions of 3N forces in nuclei near the proton drip line.

In order to learn more about the decay characteristic and ground-state mass of $^{22}$Si, more experiments with high statistics are required to be performed in the future, such as mass measurements with the storage ring and decay studies with Time Projection Chambers. Furthermore, the nature of $\beta$-delayed 2p decay via the IAS of $^{22}$Al can be investigated by using a more advanced high-granularity and large-coverage silicon array.

We acknowledge the continuous effort of HIRFL operators for providing good-quality beams and ensuring compatibility of the electronics. This work was supported by the Major State Basic Research Developing Program under Grant No. 2013CB834404, the National Natural Science Foundation of China under Grant Nos. U1432246, U1432127, U1632136, 11375268, 11305272, and 11475263, the Talented Young Scientist Program from Ministry of Science and Technology of China, the China Postdoctoral Science Foundation under Grant No. 2014M562481, the Special Program for Applied Research on Super Computation of the NSFC-Guangdong Joint Fund (the second phase), the funding of CFT (IN2P3/CNRS, France), AP th\'eorie 2015--2016, and the Ministry of Foreign Affairs and International Development of France in the framework of PHC Xu GuangQi 2015 under project No. 34457VA.

\end{document}